\documentclass[sigconf]{acmart}

\usepackage{multirow,tabularx,caption,subcaption,bm,enumitem,kantlipsum,xcolor,hyperref,fontawesome5,balance}
\usepackage{arydshln}

\AtBeginDocument{%
  }

\copyrightyear{2024} 
\acmYear{2024} 
\setcopyright{rightsretained} 
\acmConference[SIGIR '24]{Proceedings of the 47th International ACM SIGIR Conference on Research and Development in Information Retrieval}{July 14--18, 2024}{Washington, DC, USA}
\acmBooktitle{Proceedings of the 47th International ACM SIGIR Conference on Research and Development in Information Retrieval (SIGIR '24), July 14--18, 2024, Washington, DC, USA}
\acmDOI{10.1145/3626772.3657725}
\acmISBN{979-8-4007-0431-4/24/07}

\makeatletter
\gdef\@copyrightpermission{
  \begin{minipage}{0.3\columnwidth}
\href{https://creativecommons.org/licenses/by/4.0/}{\includegraphics[width=0.90\textwidth]{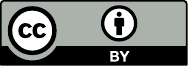}}
  \end{minipage}\hfill
  \begin{minipage}{0.7\columnwidth}
   \href{https://creativecommons.org/licenses/by/4.0/}{This work is licensed under a Creative Commons Attribution International 4.0 License.}
  \end{minipage}
  \vspace{5pt}
}
\makeatother

\acmConference[SIGIR '24]{}{July 14--18,
  2024}{Washington D.C., USA}

\begin{document}

\title[IISAN: Efficiently Adapting Multimodal Representation for Sequential Recommendation with Decoupled PEFT]{IISAN: Efficiently Adapting Multimodal Representation for Sequential Recommendation with Decoupled PEFT}

\author{Junchen Fu} 
\affiliation{
  \institution{University of Glasgow}\streetaddress{}\city{Glasgow}\country{United Kingdom}}
\email{j.fu.3@research.gla.ac.uk}
  
\author{Xuri Ge$^{\dagger}$}
\affiliation{
\institution{University of Glasgow}\streetaddress{}\city{Glasgow}\country{United Kingdom}}
\email{x.ge.2@research.gla.ac.uk}

\author{Xin Xin}
\affiliation{
\institution{Shandong University}\streetaddress{}\city{Qingdao}\country{China}}
\email{xinxin@sdu.edu.cn}

\author{Alexandros Karatzoglou}
\affiliation{
\institution{Amazon}\streetaddress{}\city{Barcelona}\country{Spain}}
\email{alexandros.karatzoglou@gmail.com}

\author{Ioannis Arapakis}
\affiliation{
\institution{Telefonica Research}\streetaddress{}\city{Barcelona}\country{Spain}}
\email{arapakis.ioannis@gmail.com}

\author{Jie Wang}\affiliation{
\institution{University of Glasgow}\streetaddress{}\city{Glasgow}\country{United Kingdom}}
\email{j.wang.9@research.gla.ac.uk}

\author{Joemon M. Jose}\affiliation{
\institution{University of Glasgow}\streetaddress{}\city{Glasgow}\country{United Kingdom}}
\email{joemon.jose@glasgow.ac.uk}

\thanks{$\dagger$ Corresponding author.}
\renewcommand{\shortauthors}{Junchen Fu et al.}

\begin{abstract}
Multimodal foundation models are transformative in sequential recommender systems, leveraging powerful representation learning capabilities. While Parameter-efficient Fine-tuning (PEFT) is commonly used to adapt foundation models for recommendation tasks, most research prioritizes parameter efficiency, often overlooking critical factors like GPU memory efficiency and training speed. Addressing this gap, our paper introduces  \textit{IISAN} (\underline{I}ntra- and \underline{I}nter-modal \underline{S}ide \underline{A}dapted \underline{N}etwork for Multimodal Representation)\footnote{Same pronunciation as the name of Thailand's largest region "Isan".}, a simple plug-and-play architecture using a Decoupled PEFT structure and exploiting both intra- and inter-modal adaptation. 

IISAN matches the performance of full fine-tuning (FFT) and state-of-the-art PEFT. More importantly, it significantly reduces GPU memory usage — from 47GB to just 3GB for multimodal sequential recommendation tasks. 
Additionally, it accelerates training time per epoch from 443s to 22s
compared to FFT. This is also a notable improvement over the Adapter and LoRA, which require 37-39 GB GPU memory and 350-380 seconds per epoch for training. 

Furthermore, we propose a new composite efficiency metric, TPME (Training-time, Parameter, and GPU Memory Efficiency) to alleviate the prevalent misconception that "parameter efficiency represents overall efficiency". TPME provides more comprehensive insights into practical efficiency comparisons between different methods. Besides, we give an accessible efficiency analysis of all PEFT and FFT approaches, which demonstrate the superiority of IISAN. Code is available at \url{https://github.com/GAIR-Lab/IISAN}.
\end{abstract}




\maketitle

\section{Introduction}
\label{sec:intro}

Large foundation models such as GPT-4 \cite{achiam2023gpt}, DALL-E \cite{ramesh2022hierarchical}, LLaMA \cite{touvron2023llama}, and CLIP \cite{radford2021learning}, are at the forefront of AI innovation, captivating the entire AI community. These models have become pivotal in AI research areas such as Natural Language Processing, Computer Vision, and Multimodal Learning Tasks. 
%%%
In particular, their superior ability to generate universal representations is highly advantageous for sequential recommendation tasks that have recently shifted from traditional reliance on IDs (Identifiers)~\cite{he2017neural,hidasi2015session} to multimodal item content (Text, Images, etc.)

While this paradigm shift is undeniably successful, it introduces significant challenges. Large foundation models, characterized by their extensive parameterization, incur substantial costs when applying traditional full fine-tuning (FFT) methods \cite{devlin2018bert}. This partly explains why much research focuses on text modality \cite{wu2020learning,wu2021empowering,hou2022learning,hou2022towards} – the training cost for visual modality is even higher. For instance, vision encoders like ViT have long sequence lengths (up to 197), compared to less than 50 in typical text scenarios.
%\footnote{16x16 patch + one cls token} 
Longer sequence length can significantly increase the activations in GPU memory, which we will explain in Section \ref{sec:theory_M}. This high training cost is even more pronounced in multimodal scenarios, combining text and visual information, which significantly increases model size and data input.
Nevertheless, despite such efficiency issues, the intuitive advantage of multimodal representation lies in its ability to comprehensively integrate information, thereby enhancing overall performance. Therefore, optimizing efficiency in multimodal recommendation is of paramount importance. 

\begin{figure}
  \centering
   \includegraphics[width=0.95\linewidth]{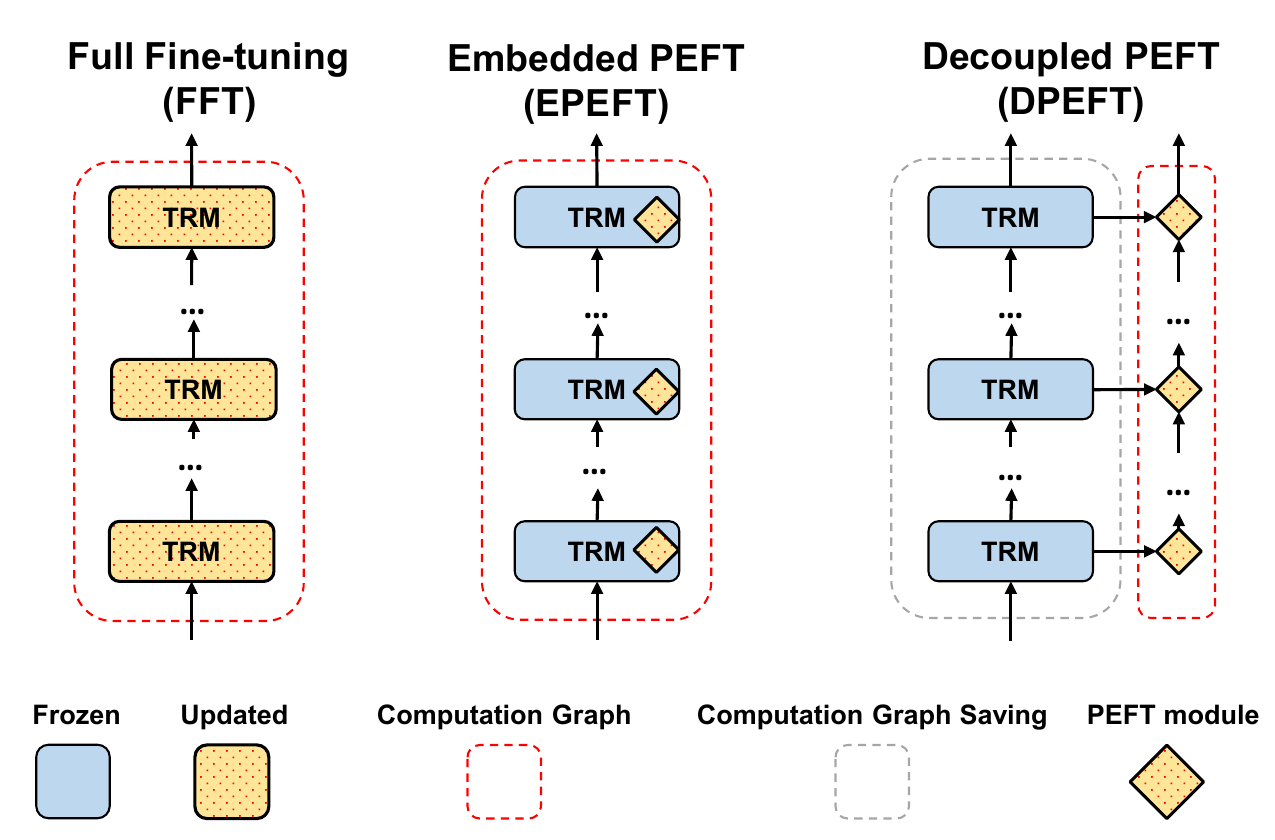}
   \vspace{-0.2in}
  \caption{Comparsions among Full Fine-tuning (FFT), Embedded PEFT and Decoupled PEFT for feature representation learning. 
  The traditional Embedded PEFT (EPEFT), e.g., Adapter \cite{houlsby2019parameter} and LoRA \cite{hu2021lora},  embed the additional trainable parameters into the foundation models, reducing trainable parameters but still having heavy computation graph during backpropagation. The proposed IISAN belongs to Decoupled PEFT (DPEFT), which significantly reduces the size of the computation graph by decoupling the PEFT from backbones and maintains the latest trainable parameters by freezing backbones.
  }
  \vspace{-0.2in}

    \label{fig:compare} 
\end{figure}

We identified two additional key challenges in the costly training of large foundation models: \textbf{GPU memory} and \textbf{training time}. The substantial expense of GPU memory, such as that required for advanced GPUs like A100 and H100 80G, poses a barrier for researchers and engineers interested in developing large foundation models but lack access to these resources. Moreover, the issue of training time encompasses not only extended waiting periods but also the escalation of electricity expenses. Crucially, the energy consumed during lengthy training processes directly contributes to higher carbon emissions, a matter of significant global environmental concern~\cite{chen2022strategies,williams2021carbon}.

In response to the FFT's efficiency problem, many researchers have turned to parameter-efficient fine-tuning (PEFT) methods, such as LoRA \cite{hu2021lora}, Adapter \cite{houlsby2019parameter} and Bitfit \cite{zaken2021bitfit}, etc.
Note that, adapter-based approaches have shown performance comparable to FFT in sequential RS, as discussed in \cite{fu2024exploring}. However, a critical question arises: \textbf{Does the parameter efficiency represent the practical model efficiency, i.e. lower parameter $\rightarrow$ lower GPU memory \& faster training speed?} %\textit{In fact, initially proposed to reduce on-disk memory , sole parameter efficiency may not address the practical concerns of researchers.}  
PEFT was initially proposed to reduce trainable parameters \cite{houlsby2019parameter}, aiming to save storage by reducing the need for multiple copies of foundation models for various tasks \cite{houlsby2019parameter, fu2024exploring}. For instance, as shown in Figure \ref{fig:compare}, popular PEFT methods \cite{houlsby2019parameter,hu2021lora,karimi2021compacter,pfeiffer2020adapterhub} embed new trainable parameters into the foundation models, namely Embedded PEFT (EPEFT). However, the computation graph of these EPEFTs, such as Adapter and LoRA, shows no significant reduction \cite{sung2022lst,cai2020tinytl}. Consequently, the cost of backward propagation, i.e. GPU memory and training time, does not decrease appreciably, which continues to be a primary bottleneck during the training stage, as detailed in Section \ref{sec:theory}.

As mentioned before, we argue that two intuitive key issues exist in PEFT research: 
\begin{itemize}
\item Current embedded PEFT methods 
are inherently flawed, i.e., the heavy computation graph in backpropagation. 
\item In the PEFT research methodologies, there is a misconception that parameter efficiency directly translates to overall efficiency, and the model efficiency evaluation (only focusing on the size of trainable parameters) is lopsided.
\end{itemize}
Current research in multimodal sequential recommendation is suffering from these critical issues \cite{fu2024exploring,geng2023vip5,cui2022m6}. 
Motivated by this, we conduct our research. 
\textcolor{black}{Specifically, we develop a \textit{a simple yet efficient} Intra- and Inter-modal Side Adapted Network (IISAN) to adapt the multimodal foundation models more efficiently. Compared to traditional EPEFT (Adapter and LoRA, etc.), IISAN's innovations encompass three aspects: (1) We adopt the advanced Decoupled PEFT (DPEFT) structure, described in Figure \ref{fig:compare}, to largely reduce the computation graph; (2) Drawing from our insights into the characteristics of the DPEFT, we further propose to improve efficiency by implementing a caching strategy, as detailed in Section \ref{sec:iisan}; (3) Based on the traits of multimodality, we embrace the capabilities of both unimodality (intra-SAN) and multimodality interactions (inter-SAN) for better multimodal representation learning.} 

 Additionally, we correct the misconception that "parameter efficiency is equal to model practical efficiency" in two aspects.
First, we undertake a thorough yet accessible analysis in Section \ref{sec:theory}. This analysis focuses on the practical efficiency of various prevalent approaches, including full fine-tuning (FFT), EPEFTs (such as LoRA \cite{hu2021lora} and Adapter \cite{houlsby2019parameter}, etc.), and DPEFT (IISAN). The evaluation encompasses three key facets: training speed, trainable parameters,  and GPU memory usage. Second, to unify the point of evaluating the practical model efficiency, we introduce a metric named TPME (Training-time, Parameter, and GPU Memory Efficiency), which comprehensively considers the above three critical factors to address the second aforementioned issue in PEFT research.

Overall, the major contributions of our paper are threefold:
\begin{itemize}
\item We introduce a novel Intra- and Inter-modal Side Adapted Network (IISAN) following the decoupled PEFT paradigm for end-to-end sequential recommendation tasks based on pre-trained multimodal foundation models. IISAN is allowed to employ caching strategy to further improve efficiency due to its DPEFT property.
\item We propose a new practical efficiency metric (TPME), harmonizing the costs of training time, trainable parameters, and GPU memory, which offers a holistic measure of model efficiency.
\item We provide a detailed and accessible analysis for understanding the high efficiency of the PEFT mechanism. It positions the IISAN as a theoretically superior architecture in terms of practical efficiency compared to prevalent PEFT.
\end{itemize}
Extensive experiments validate that IISAN achieves comparable performance over the FFT and the state-of-the-art PEFT methods on three widely-used multimodal recommendation datasets, while also delivering a notable enhancement in GPU memory and training time efficiency. 

\begin{figure*}
  \centering
\includegraphics[width=0.9\linewidth]{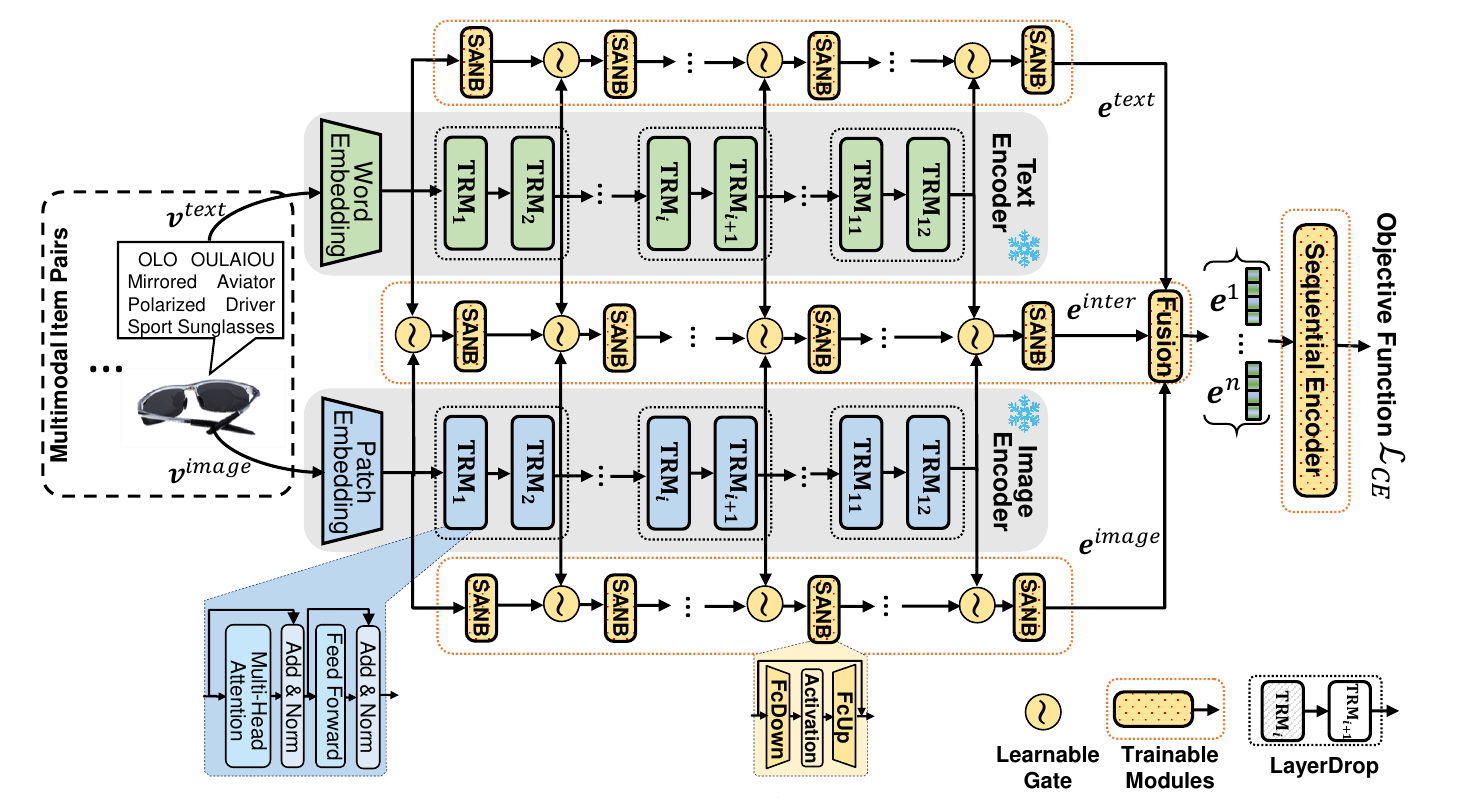}
   \vspace{-0.2in}
  \caption{An Overview of the IISAN for sequential recommendation. The framework takes the pre-trained text encoder BERT \cite{devlin2018bert} and image encoder ViT \cite{dosovitskiy2020image} as an example, which contains 12 Transformer-blocks (TRMs) respectively. IISAN proposes intra- and inter-modal side adapted networks, where the intra-modal SANs mainly construct independent adaptive representation learning within two modalities and the inter-modal SAN focuses on the efficient multimodal interactions between layer hidden states in multimodal networks. SANs consist of multiple SAN blocks (SANBs) and learnable fusion gates. Each SANB receives the hidden states from the corresponding layers and makes an adaptive learning optimization for the final recommendation task by a unified objective function. Notablely, we leverage LayerDrop to further omit redundancy.}
    \label{fig:iisan} 
      \vspace{-0.1in}
\end{figure*}
\begin{figure}
  \centering
   \includegraphics[width=0.9\linewidth]{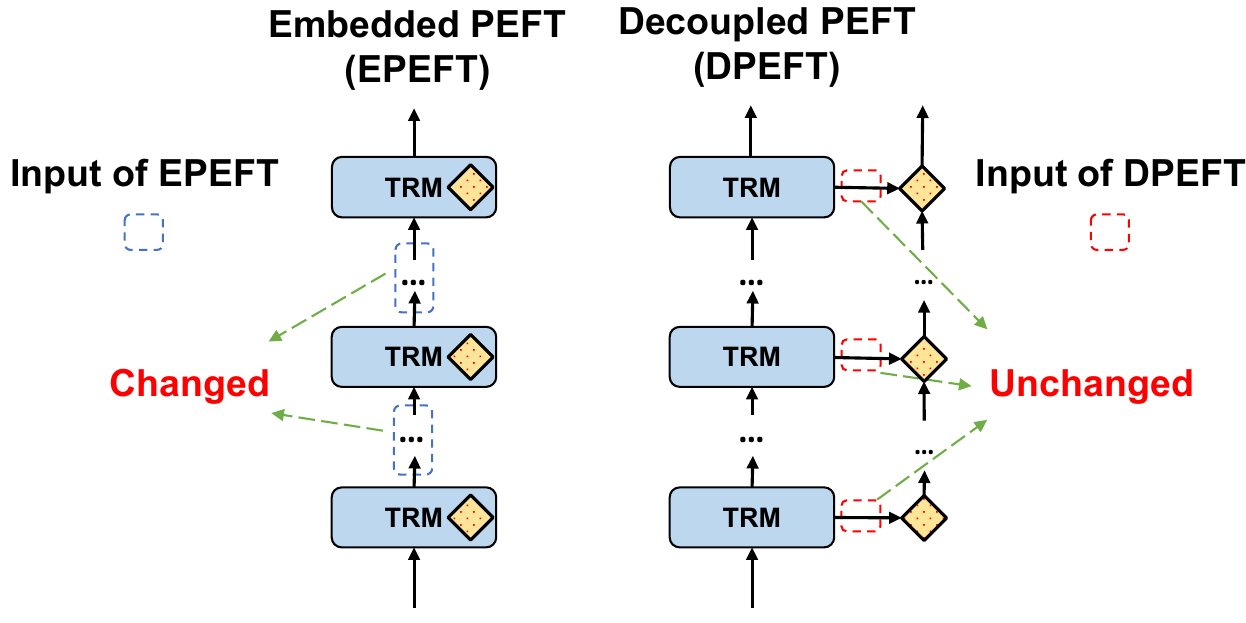}
   \vspace{-0.2in}
  \caption{Caching strategies comparison. The input for the DPEFT remains constant and, in theory, can be cached. On the other hand, the input for the EPEFT is subject to change as it is influenced by parameter updates in the last block. 
  }
    \label{fig:caching} 
    \vspace{-0.1in}
\end{figure}

\section{Methodology}

 The overall framework of the proposed IISAN is shown in Figure \ref{fig:iisan}. 
 We depart from the mainstream multimodal representation learning models \cite{yuan2023go,fu2024exploring,wang2022transrec,ge2021structured,wu2020learning,wu2021empowering} to introduce a novel high-efficiency paradigm for personalized fine-tuning in sequential recommendation—referred to as decoupled parameter-efficient fine-tuning (decoupled PEFT, DPEFT).
This innovative approach specifically addresses a critical aspect: the efficient adaptation of pre-trained large-scale multimodal models as item encoders. Initially, we propose to decouple the new trainable side adapted networks (SAN) and the frozen multimodal backbones. This decoupling is designed to optimize the extensive computation graph of backpropagation, effectively tackling challenges in training time and GPU memory encountered when transitioning large-scale models to downstream tasks. Leveraging the unique features of DPEFT, we introduce caching strategies for IISAN to significantly enhance its practical efficiency.
Furthermore, capitalizing on the intrinsic nature of multimodal representations, the proposed IISAN architecture introduces two independent intra-modal SANs for visual and textual modality adaptation, along with an inter-modal SAN for handling interactions between the two modalities. The decoupled intra- and inter-modal SANs adeptly transfer pre-trained large-scale multimodal fundamental models, such as BERT \cite{devlin2018bert}, DEBERTA \cite{he2021debertav3,he2020deberta}, ViT \cite{dosovitskiy2020image}, and CLIP \cite{radford2021learning}, to downstream multimodal recommendation tasks while maintaining multifaceted efficiency.
Moreover, we pioneer a new efficiency metric—TPME—to evaluate the practical training efficiency of models, diverging from a reliance solely on trainable parameters.
\vspace{-0.1in}

\subsection{Our method: IISAN} \label{sec:iisan}
Given a recommendation dataset $\mathcal {D} = \{\mathcal {U}, \mathcal {V}\}$ where $\mathcal {U}$, $\mathcal {V}$ denote the set of users and the set of items, respectively. For a multimodal sequential recommendation task, we aim to predict the next item interacted with user $u$ by exploiting his/her $n$ past behaviors. In multimodal recommendation, each item $v$ contains two modal representations, i.e. text  ($v^{text}$) and corresponding image ($v^{image}$).  
We feed the texts and images of items into two pre-trained textual and visual fundamental models, such as BERT \cite{devlin2018bert} for texts and ViT \cite{dosovitskiy2020image} for images, which consist of an embedding layer (word embedding for texts and patch embedding for images) and multiple transformer blocks respectively. 
Through the pre-trained multimodal backbones, we can obtain the textual and visual embeddings from the embedding layers and the multiple hidden states ($\{h_i^{text}\}$ and $\{h_i^{image}\}$) from the transformer blocks ($\{TRM_i\}$), where $i$ denotes the $i$-th layer of the backbone.

In this paper, our model innovation lies in that we propose intra- and inter-modal side adapted network (IISAN) aims to maximize the utilization of knowledge derived from pre-trained models in understanding multimodal item sequence representation by a decoupled PEFT paradigm. 
Specifically, we decouple the trainable parameters into three separate towers, which are the textual-modality side adapted network for text representation training, the visual-modality side adapted network for image representation training and an inter-modal side adapted network for image-text interactive representation training. 
Similar to Adapter \cite{houlsby2019parameter}, each side adapted network (SAN) consists of multiple SAN blocks (SANBs), each of which contains Upsample layers and downsample layers based on fully-connected network.\footnote{SANB, a network unit of SAN, utilizes the network block similar to Adapter~\cite{houlsby2019parameter}. However, unlike the Adapter's embedded approach, SANB is implemented in a decoupled way.}
As shown in Figure \ref{fig:iisan}, the structure of textual SAN, visual SAN and inter-SAN have structure symmetry, due to the consistency of the backbone models. 
Taking the textual SAN as an example, different from complex fusion ways \cite{ge2019colloquial}, a learnable gate mechanism is employed to fusion the information $\{h_{i-1}^{B^{intra}}\}$ from the last SANB and the current hidden state $\{h_i^{text}\}$, as the formula:
\begin{equation}
    h_i^{B^{intra}} = SANB^{intra}\left(\mu_i^{text} \ast h_{i-1}^{B^{intra}} + (1 - \mu_i^{text}) \ast h_i^{text}\right)
\end{equation}
where $\mu_i^{text}\in [0,1]$. Note that the first SANB only inputs the text embeddings and the visual SAN employs similar operations. 
For the inter-SAN, we employ a similar gating method.
%\textcolor{red}{some text is missing at this point ...}. 
We design a fusing mechanism to  fuse the hidden states of two modalities and sum the information from the last SANB, as the formula:
\begin{equation}
      h_i^{B^{inter}} = SANB^{inter}\left(\beta_i \ast h_i^{image} + (1 - \beta_i) \ast h_i^{text} + h_{i-1}^{B^{inter}}\right)
\end{equation}
where $\beta_i\in [0,1]$.  Note that the first inter-SANB only inputs the text embedding and the visual embeddings. 

Additionally, to further enhance network efficiency and address issues of layer redundancy, we introduce a LayerDrop techniques, like \cite{sung2022lst,fan2019reducing}, to save the number of SAN blocks. Specifically, we group two transformer blocks together and drop the first hidden state to the SANs, which can save half the amount of SANBs. 
Moreover, we also explore different LayerDrop schemes in Section \ref{sec:rq4}, i.e. dropping different hidden layer states, to achieve the best balance between efficiency and effect, which reflects the significance of different encoder layers for the final modality representation.

Finally, we obtain an efficient new multimodal item representations from the  intra- and inter-modal SANs, including the textual representations $\{e^{text}\}$, the visual representations $\{e^{image}\}$ and the textual-visual interacted representations $\{e^{inter}\}$. 
A linear-based fusion layer $(FL)$ is added to ensure the consistency of the output dimensions of the item embedding and input dimensions of the sequential encoder for the final recommendation, as follows: 
\begin{equation}
{e^{item}} = FL([{e^{image}} : {e^{inter}}:{e^{text}}])
\end{equation}
where $[:]$  means the feature concatenation. Then, we input the $\boldsymbol{e^{item}}$ into the sequential encoders and calculate the final predicted score for user $u$ to $i$-th item as $\boldsymbol{\hat{y}_{ui}}$, which is the product of the output sequential encoder and corresponding item embedding. 

In terms of training details, we exploit the in-batch debiased Cross-Entropy loss function $\mathcal{L}_{CE}$~\cite{yi2019sampling,yuan2023go,ji2023online} widely adopted in both academic literature and industrial systems.
\begin{equation}
D_{ui} = \exp(\hat{y}_{ui} - \log(p_i)) + \sum_{j \in [B], j \notin I_u} \exp(\hat{y}_{uj} - \log(p_j))
\end{equation}
\begin{equation}
\mathcal{L}_{CE}=-\sum_{u \in \mathcal{U}} \sum_{i \in [2,...,n+1]} \log \frac{\exp(\hat{y}_{ui} - \log(p_i))}{D_{ui}}
\end{equation}
where  $p$ is the popularity of the item, $I_u$ and $B$ stand for the item set interacted by user $u$ and the batch. $n+1$ item denotes the predicted item for user $u$.  $D_{ui}$ is a temporary variable to simplify Formula (5).

Introducing an innovative approach to enhance efficiency, we suggest a caching technique as a refinement strategy \cite{karedla1994caching,chen2020mobile}. As shown in Figure \ref{fig:caching}, due to the advantages of the decoupled PEFT mechanism, i.e., the separable  pre-trained backbone model and the new trainable model, made this possible.  This technique entails the storage and reuse of item hidden states extracted from pre-trained multimodal backbones, minimizing the necessity for repeated forward passes of foundational models during training. Notably, it is crucial to highlight that this approach may not be applicable to Embedded PEFT methods.  Because the input for each PEFT module in Embedded PEFTs will be altered by the previous module, which leads the layer’s hidden states to change and is unsuitable for caching. This limitation of EPEFT  highlights the Decoupled PEFT's superior efficiency.

\subsection{New Efficiency metric: TPME}
In this paper, we propose a new composite metric\footnote{The composite metric, frequently used in social statistics, measures levels across multiple dimensions. The well-known Human Development Index (HDI) \cite{sagar1998human}, employed by the United Nations, incorporates dimensions of education, longevity, and income. Our TPME follows the HDI's fundamental calculation principle, integrating various dimensions for a comprehensive evaluation.}  (termed TPME) to evaluate the practical efficiency of different PEFT methods, e.g. Adapter, LoRA, etc. It integrates the efficiencies of training time, trainable parameters, and GPU memory into a unified evaluation metric by adjustable practical factors.
Specifically, assuming we evaluate the efficiency of $i$-th model among $K$ models, TPME is calculated based on their training time $T =\{t_1,...,t_K\}$, trainable parameters $P =\{p_1,...,p_K\}$ and GPU memory  $M =\{m_1,...,m_K\}$  as follows:
\begin{equation}
t_i^{\text{norm}} = \frac{t_i - \min T}{\max T - \min T}
\end{equation}
\begin{equation}
p_i^{\text{norm}} = \frac{p_i - \min P}{\max P - \min P}
\end{equation}
\begin{equation}
m_i^{\text{norm}} = \frac{m_i - \min M}{\max M - \min M}
\end{equation}
\begin{equation}
TPME_i = \alpha_1 t_i^{\text{norm}}  + \alpha_2 p_i^{\text{norm}} + \alpha_3 m_i^{\text{norm}}
\end{equation}
\begin{equation}
\alpha_1 + \alpha_2 + \alpha_3 = 1
\end{equation}
where $\alpha$ denotes the weighting assigned to each term, tailored to specific circumstances.
For example, in scenarios where only a limited GPU capacity is available for model training, it's advisable to significantly augment the weight of $M$. Within the scope of this paper, we've adjusted the values of $\alpha_1$ and $\alpha_3$ to 0.45, and $\alpha_2$ to 0.1. This adjustment reflects our focus on two key practical aspects: training speed and memory efficiency. It is crucial to understand that the Training-time, Parameter, and GPU Memory Efficiency (TPME) framework is specifically intended for the comparative analysis of a minimum of two methodologies. For example, this could involve a comparison between methodologies such as FFT and IISAN. 
%\textcolor{red}{methodologies or parameters?}. 
For each individual factor, it's crucial to ensure consistency in value derivation by conducting evaluations within the same experimental environment and setup, thus preserving the integrity and relevance of the results. An example of calculation is provided in Section \ref{sec:rq1}.

\begin{table*}
  \caption{Efficiency Comparison of FFT and PEFT. In the Training-time metric, the $FP/fp$, $BP/bp$, and $WU/wu$ represent the training time of the forward pass, backward propagation, and weight update respectively. Within the Parameter metric, $TP/tp$ symbolizes the model of trainable parameters. Concerning the GPU memory metric, $MP/mp$ and $A/a$ denote the model parameters and activations, respectively. Notably, in every instance, the lower case variables (e.g., $fp$, $bp$, $wu$) are significantly less than ($\ll$) their upper case counterparts (e.g., $FP$, $BP$, $WU$). We utilize "$>$" and "$=$" to denote relationships of greater and equal magnitude, respectively. } 
  \vspace{-0.1in}
  \renewcommand\tabcolsep{5pt}
  \renewcommand{\arraystretch}{0.9}
  \label{tab:eff}
  \begin{tabular}{ccccccccccc}
    \hline
    \multirow{2}{*}{Metric}&\multirow{2}{*}{FFT} &\multirow{2}{*}{ }&\multirow{2}{*}{Adapter}&\multirow{2}{*}{ }&\multirow{2}{*}{LoRA}&\multirow{2}{*}{ }&\multirow{2}{*}{IISAN (Uncached)}&\multirow{2}{*}{ }&\multirow{2}{*}{IISAN (Cached)}\\
    &&&&\\
    \hline
    Training-time&$O(FP+BP+WU)$ &$>$&$O(FP+BP+wu)$&$=$&$O(FP+BP+wu)$&$>$&$O(FP+bp+wu)$&$>$&$O(fp+bp+wu)$\\
    Parameter&$O(TP)$&$>$&$O(tp)$&$=$&$O(tp)$&$=$&$O(tp)$&$=$&$O(tp)$\\
    GPU Memory&$O(MW+A)$&$=$&$O(MW+A)$&$=$&$O(MW+A)$&$>$&$O(MW+a)$&$>$&$O(mw+a)$\\

  \hline
\end{tabular}
\vspace{-0.15in}
\end{table*}

\section{Analysis of Efficiency} \label{sec:theory}

This section analyzes the detailed composite efficiency of training downstream  
 tasks, such as recommender systems and cross-modal retrieval, etc., with PEFTs based on pre-trained large-scale multimodal foundation models to compensate for filling in the absences in the existing literature\footnote{Note that we only study from the model perspectives, other engineering approaches such as Quantization applied in QLoRA \cite{dettmers2023qlora} is not in our scope.}. 
 While quantifying exact values in Transformer-based architectures is complex, we provide approximate values with an upper bound $O$, facilitating comparative analysis with various approaches in Table \ref{tab:eff}. In this comparison, we keep different dimensional variables inside $O$ and diminish the smaller variables within the same dimension and constant coefficients using "$\approx$", following the algorithm's fundamental principles \cite{skiena1998algorithm}.

\subsection{Training-time Efficiency}
Inspired by \cite{larochelle2009exploring,safayenikoo2021weight}, we analyze the time spent on three key components, i.e. forward passes ($FP$), backward passes ($BP$) and weight updates ($WU$), of one training iteration.
First, we define the aforementioned three components for the large-scale foundation model as $O(FP)$, $O(BP)$ and $O(WU)$, and for the smaller PEFT network as $O(fp)$, $O(bp)$, and $O(wu)$, where $FP$$\gg$$fp$, $BP$$\gg$$bp$ and $WU$$\gg$$wu$. 
Consequently, the total training time of one iteration in full fine-tuning (FFT) is around $O(FP+BP+WU)$.
For LoRA \cite{hu2021lora} and Adapter \cite{houlsby2019parameter}, their forward and backward pass involves both foundation models and PEFT components. In addition, since the foundation model is frozen, the weight update time includes only the PEFT model $O(wu)$. So the total training time for LoRA/Adapter is:

\begin{equation}
    O(FP+fp+BP+bp+wu) \approx O(FP+BP+wu)
\end{equation}
However, for  IISAN, due to its decoupled structure, we can omit the backward propagation through the large foundation models. So the total training time is:
\begin{equation}
    % O(FP+fp+bp+wu) \Leftrightarrow O(FP+bp+wu)
    O(FP+fp+bp+wu) \approx O(FP+bp+wu)
\end{equation}
Moreover,  as discussed in Section \ref{sec:iisan},  IISAN (Cached) approach can further save the forward pass time of the foundation models, i.e., training time is $O(fp+bp+wu)$.

\subsection{Parameter Efficiency}
\label{sec:theory_P}
The parameter efficiency is considered as a determinant of model efficiency in many research papers \cite{fu2024exploring,houlsby2019parameter}. The more trainable parameters a method has, the fewer parameter efficient it becomes. Therefore, we analyze the entire backbone models with $O(TP)$ parameter sizes when FFT, while the PEFT components (including Adapter, LoRA, and IISAN, etc.) only contain the negligible $O(tp)$ trainable parameters, where $tp \ll TP$. 
As mentioned in \cite{fu2024exploring,houlsby2019parameter}, the main reason for the parameter efficiency being a concern is due to storage inefficiencies during downstream task model migration, especially on mobile. However, with the popularity of cloud-based storage, we believe that this should not be the only focus for recommendation model training, the training time and GPU memory efficiencies are more important in practice.

\subsection{GPU Memory Efficiency}
\label{sec:theory_M}
 Different from existing methods, a detailed analysis of GPU Memory efficiency --on-GPU memory composite-- is given to explain why traditional PEFTs cannot reduce much GPU memory, while IISAN can.

The consumption of GPU memory is primarily distributed across five aspects\footnote{The Anatomy of Model's Memory Section from \url{https://huggingface.co/docs/transformers/v4.18.0/en/performance}}, i.e. (i) model weights, (ii) gradients, (iii) optimizer states, (iv) forward activations saved for gradient computation, and (vi) Others (temporary buffers, functionality-specific memory, etc.).
The main sources of GPU memory usage are the first four components.
The fifth component is usually a relatively small one that is  omitted here. 

In particular, we first equate the model's gradients with its trainable weights\footnote{Gradients normally equal to 4 bytes * number of trainable model weights, we denote them as equal value for simplicity}, denoting them as $O(MW)$ for the backbone model and $O(mw)$ for the PEFT modules, where $MW$$\gg$$mw$. Secondly, the optimizer states take the standard Adam optimizer \cite{kingma2014adam} as an example.
It doubles the parameter count for calculating first and second momentum orders, so the optimizer states are $O(2MW)$ for backbone models and $O(2mw)$ for PEFT modules. %Note that activations depend on the computation graph's size. \footnote{Activations are also affected by sequence length, hidden size, and batch size \cite{gomez2017reversible,gao2021scaling, cai2020tinytl}.} 
Additionally, we denote the activations for the backbone models and the PEFT modules by $O(A)$ and $O(a)$ respectively, where $A \gg a$. As stated in \cite{gomez2017reversible,gao2021scaling, cai2020tinytl}, the activations mainly depend on the computation graph's size\footnote{Activations are also influenced by sequence length, hidden size and batch size, etc}. This is because backpropagation requires storing the activations of the network in GPU memory, the memory cost that is proportional to the number of units in the network. 
Finally, we can represent the total GPU memory of FFT by:
 \begin{equation}
    O(MW + MW + 2MW + A)  = O(4MW+A) \approx O(MW+A)
    \label{eq:UE}
\end{equation}
For the LoRA/Adapter-based approach, they reduce the trainable parameters from $O(MW)$ to $O(mw)$, but the computation graph as shown in Figure \ref{fig:compare}, has not been reduced. Therefore, the total GPU memory for LoRA/Adapter is:
 \begin{equation}
    O(MW+mw+mw+2mw+A+a) =  O(MW+4mw+A+a)\approx O(MW+A)
    \label{eq:UE}
\end{equation}

Note that while the LoRA and adapter share the same theoretical complexity as the FFT, they can still save up to three times the GPU memory when the model parameters are the bottleneck\footnote{Typically, GPU memory usage is influenced by either model size or activations. In the context of Multimodality-based sequential recommendation scenarios, where sequences tend to be lengthy (item token length $*$ sequence length $*$ batch size $*$ types of modalities), activations often become the bottleneck. Consequently, IISAN demonstrates significant efficiency in the multimodal recommendation, particularly in scenarios characterized by extensive activations.}, as highlighted in \cite{hu2021lora}. On the other hand, IISAN can further save the computation graph of the backbone model. Consequently, the total GPU memory required for IISAN would be:
 \begin{equation}
O(MW +mw + mw + 2mw + a) = O(MW+4mw+a) \approx O(MW+a)
    \label{eq:UE}
\end{equation}
Furthermore, applying the caching strategies, the IISAN (cached) alleviates memory constraints for storing foundation model weights. Consequently, the GPU memory allocation for IISAN (Cached) is as follows:
 \begin{equation}
O(mw + mw + 2mw + a ) = O(4mw + a ) \approx O(mw+a)
    \label{eq:UE}
\end{equation}

\begin{table}
  \caption{Dataset Description}
  \vspace{-0.1in}
  \renewcommand\tabcolsep{5.3pt}
  \renewcommand{\arraystretch}{0.7}
  \label{tab:dataset}
  \begin{tabular}{clcccc}
    \hline
    \multirow{2}{*}{Dataset}&\multirow{2}{*}{Users}&\multirow{2}{*}{Items}&\multirow{2}{*}{Interaction}&\multirow{2}{*}{Content}\\
    &&&&&\\
    \hline
    Scientific&12,076&20,314&81,711&Text\&Image\\

    Instrument&10,000&19,246&75,022&Text\&Image\\
    Office&10,000&22,785&71,276&Text\&Image\\
    
  \hline
\end{tabular}
\vspace{-0.1in}
\end{table}

\begin{table*}
\caption{Comparisons of efficiency-performance balance. Two key categories of evaluation metrics are used, i.e. ``Performance" and ``Efficiency". All results of HR@10 and NDCG@10 are denoted in the percentage (\%). The 'Training', 'Parameter' and 'GPU Memory' denote training time (seconds for each epoch), trainable parameters and max GPU memory usage, respectively. ``Relative Improvement" means the result differential between IISAN and FFT. Larger values ($\uparrow$) of the Performance metric indicate better performance, while smaller values ($\downarrow$) of the Efficiency metric reflect better efficiency.  ``*” denotes that the improvements of IISAN compared with FFT are significant at the level of 0.05 with paired T-test. To clearly represent the difference between uncached IISAN and cached IISAN (in blue), we divided them into two separate columns. The best results are in bold.}
\vspace{-0.1in}
\centering
\renewcommand\tabcolsep{2.5pt}
\renewcommand{\arraystretch}{0.9}
\label{tab:eff_per}
\begin{tabular}{c|cc|cccccc|cc}
\hline
\multirow{2}{*}{\textbf{Dataset}} & \multirow{2}{*}{\textbf{Category}} & \multirow{2}{*}{\textbf{Metric}} & \multirow{2}{*}{\textbf{FFT}}  & \multirow{2}{*}{\textbf{Adapter}\cite{houlsby2019parameter}} & \multirow{2}{*}{\textbf{LoRA}\cite{hu2021lora}} & \multirow{2}{*}{\textbf{BitFit}\cite{zaken2021bitfit}} & \multicolumn{2}{c|}{\textbf{IISAN (ours)}} & \multicolumn{2}{c}{\textbf{Relative Improvement}} \\
& & & & & & & Uncached & Cached & Uncached & Cached \\
\hline
\multirow{6}{*}{Scientific}&\multirow{2}{*}{Performance($\uparrow$)}
& HR@10 & 6.62  &6.61 &6.62 &6.37 &\multicolumn{2}{c|}{\textbf{6.83$^*$}} & \multicolumn{2}{c}{+3.07\%} \\
& & NDCG@10 & 4.06  &3.91 &4.09 &3.76&\multicolumn{2}{c|}{\textbf{4.14$^*$}} & \multicolumn{2}{c}{+1.93\%} \\
\cdashline{2-11} %\cline{2-11}
&\multirow{4}{*}{Efficiency($\downarrow$)}
& Training-time &443s &354s &378s &403s &\textbf{179s}& \textcolor{blue}{\textbf{22s}} & -59.59\% & -95.03\% \\
&& Parameter&195M &5M &0.8M &\textbf{0.4M} &4M&\textcolor{blue}{4M} & -97.89\% &-97.89\% \\
&&GPU Memory &46.76G &37.82G &39.07G &36.97G &\textbf{8.32G}& \textcolor{blue}{\textbf{3.11G}} & -82.19\%&-93.35\% \\ \cdashline{3-11} 
&& TPME &100\% &71.50\% &75.14\% &70.82\% &\textbf{22.34\%}& \textcolor{blue}{\textbf{0.19\%}} & -78.10\% & -99.80\% \\
\hline
\multirow{6}{*}{Instrument}&\multirow{2}{*}{Performance($\uparrow$)}
& HR@10 & 8.76  & 8.82 & 8.43 & 8.65 &\multicolumn{2}{c|}{\textbf{9.06$^*$}} & \multicolumn{2}{c}{+3.31\%} \\
&& NDCG@10 & 6.76 & 6.83 & 6.64 & 6.71 &\multicolumn{2}{c|}{\textbf{7.01$^*$}} & \multicolumn{2}{c}{+3.57\%} \\
\cdashline{2-11}
&\multirow{4}{*}{Efficiency($\downarrow$)}
& Training-time & 369s & 295s & 308s & 287s &\textbf{142s} &\textcolor{blue}{\textbf{18s}} &-61.52\% &-95.12\% \\
&& Parameter & 195M & 5M & 0.8M & \textbf{0.4M} &4M &\textcolor{blue}{4M} & -97.89\% & -97.89\% \\
&&GPU Memory & 46.76G & 37.82G & 39.07G & 36.97G &\textbf{8.32G} &\textcolor{blue}{\textbf{3.11G}} & -82.19\% &-93.35\% \\  \cdashline{3-11} 
&& TPME & 100\% & 71.55\% & 74.28\% & 69.40\% &\textbf{21.46\%}&\textcolor{blue}{\textbf{0.19\%}} &-78.54\% &-99.81\% \\
\hline
\multirow{6}{*}{Office}&\multirow{2}{*}{Performance($\uparrow$)}
& HR@10 & 6.30 & 6.65 & 6.55 & 6.78 &\multicolumn{2}{c|}{\textbf{6.80$^*$}} & \multicolumn{2}{c}{+7.35\%} \\
&& NDCG@10 & 4.58 & 4.85 & 4.78 & 4.87 &\multicolumn{2}{c|}{\textbf{4.92$^*$}} & \multicolumn{2}{c}{+6.91\%} \\
\cdashline{2-11}
&\multirow{4}{*}{Efficiency($\downarrow$)}
& Training-time & 355s & 296s & 308s & 288s &\textbf{140s} &\textcolor{blue}{\textbf{19s}} & -60.56\%&-94.65\% \\
&& Parameter & 195M & 5M & 0.8M & \textbf{0.4M} &4M &\textcolor{blue}{4M} & -97.89\%&-97.89\% \\
&&GPU Memory & 46.76G & 37.82G & 39.07G & 36.97G &\textbf{8.32G}&\textcolor{blue}{\textbf{3.11G}} & -82.19\% &-93.35\% \\  \cdashline{3-11} 
&& TPME & 100\% & 73.12\% & 75.80\% & 70.93\% &\textbf{21.77\%}&\textcolor{blue}{\textbf{0.19\%}} & -78.23\% &-99.81\% \\
\hline
\end{tabular}
\vspace{-0.1in}
\end{table*}

\section{Experiment Setup}
\textbf{Datasets.} In this paper, each item for a user sequence is represented by its raw image and text modality. We perform the preprocessing following  \cite{yuan2023go,fu2024exploring,li2023exploring}. To evaluate our methods on items with both raw image and text modality, we adopt the Amazon review datasets \cite{ni2019justifying}. We adopt three widely used datasets from it, “Industrial and
Scientific”, “Musical Instruments”, and
“Office Products”.\footnote{We only keep the items with both images and text information. Due to the massive amount of data for "Office" and "Instrument", we randomly sampled 10,000 users from the remaining datasets to align with "Scientific" following \cite{fu2024exploring}.} Due to the constraint of GPU memory, we follow \cite{fu2024exploring,yuan2023go} and set the sequence length to 10. After the preprocessing, the details of the datasets are shown in Table~\ref{tab:dataset}.

\noindent\textbf{Performance Evaluations.}
 Drawing on prior studies \cite{he2017neural, yuan2023go}, our approach involves dividing the datasets using a leave-one-out method. This involves using the final item in the interaction sequence for testing, the item before the last is used for validation, and the remaining items for training purposes. For assessment, we use HR@10 (Hit Ratio) and NDCG@10 (Normalized Discounted Cumulative Gain) \cite{hou2022learning,yuan2023go} as the primary metrics. Unless otherwise specified, all reported results pertain to the test set. Additionally, it's important to note that the predicted item is compared against the entire set of items \cite{krichene2020sampled}.

 \noindent\textbf{Implementation Details.} 
 The "bert-base-uncased", "vit-base-patch16-224", "clip-vit-base-patch16", and "deberta-v3-base" from the Huggingface platform \cite{jain2022hugging} are used as the text and image encoders in this research, respectively. The dimension of hidden representations of the sequential encoder is searched in \{32,64,128\} and set to 64. The number of the sequential encoders' Transformer blocks and attention heads is fixed to 2. We apply Adam as the optimizer without weight decay throughout the experiments and extensively search the learning rate from 1e-5 to 1e-3 while keeping the dropout probability at 0.1.
 %\textcolor{red}{the following sentence incomplete}
We search the batch size in \{8, 16, 32, 64, 128\} and obtain the largest optimal batch size to maximize GPU memory. The hidden dimension of the adapter networks and LoRA rank are carefully searched in \{8, 16, 32, 64, 128\}, respectively.
 All hyper-parameters are determined according to the performance in the validation data. All results are reported in the test set. We perform all the experiments on an A6000 GPU.

\section{Experiment}
\begin{itemize}
    \item \textbf{RQ1:} How does the performance of the proposed IISAN compare to FFT and existing PEFT methods?
    Can IISAN achieve significant efficiency without sacrificing performance?   
    \item \textbf{RQ2:} How robust is IISAN on different multimodal backbones?
    \item \textbf{RQ3:} How does  components of the proposed IISAN affect the recommendation performance and efficiency, including the LayerDrop, modality selection, gated fusion, and SANB implementation? 
    \item \textbf{RQ4:} This proposed IISAN explores multimodal scenarios; does it have any advantages over unimodal methods (Text-only and Image-only)?
\end{itemize}

\subsection{Efficiency-Performance Balance (RQ1)}
\label{sec:rq1}
We perform extensive experiments on three multimodal sequential recommendation datasets and compare IISAN with full fine-tuning (FFT) based on \cite{fu2024exploring} in terms of model efficiency and performance. Furthermore, we also include the state-of-the-art PEFT approaches, such as Adapter \cite{houlsby2019parameter}, LoRA \cite{hu2021lora}, and BitFit \cite{zaken2021bitfit}. The Adapter \cite{houlsby2019parameter} adopts the classic Houlsby architecture which has been proven to be the most effective PEFT approach for unimodality-based recommendation in \cite{fu2024exploring}. The LoRA \cite{hu2021lora} is widely used due to its smaller trainable parameters. BitFit, a common baseline for PEFT-based recommendation research \cite{hu2021lora,sung2022lst,karimi2021compacter}, updates only the bias layers within the entire network to achieve the minimum trainable parameters. Notable, the current mainstream PEFT method in the recommendation is embedded PEFT (EPEFT) \cite{houlsby2019parameter,hu2021lora,zaken2021bitfit}, and the proposed IISAN belongs to the decoupled PEFT (DPEFT).

\begin{figure}[t]
  \centering
   \includegraphics[width=\linewidth]{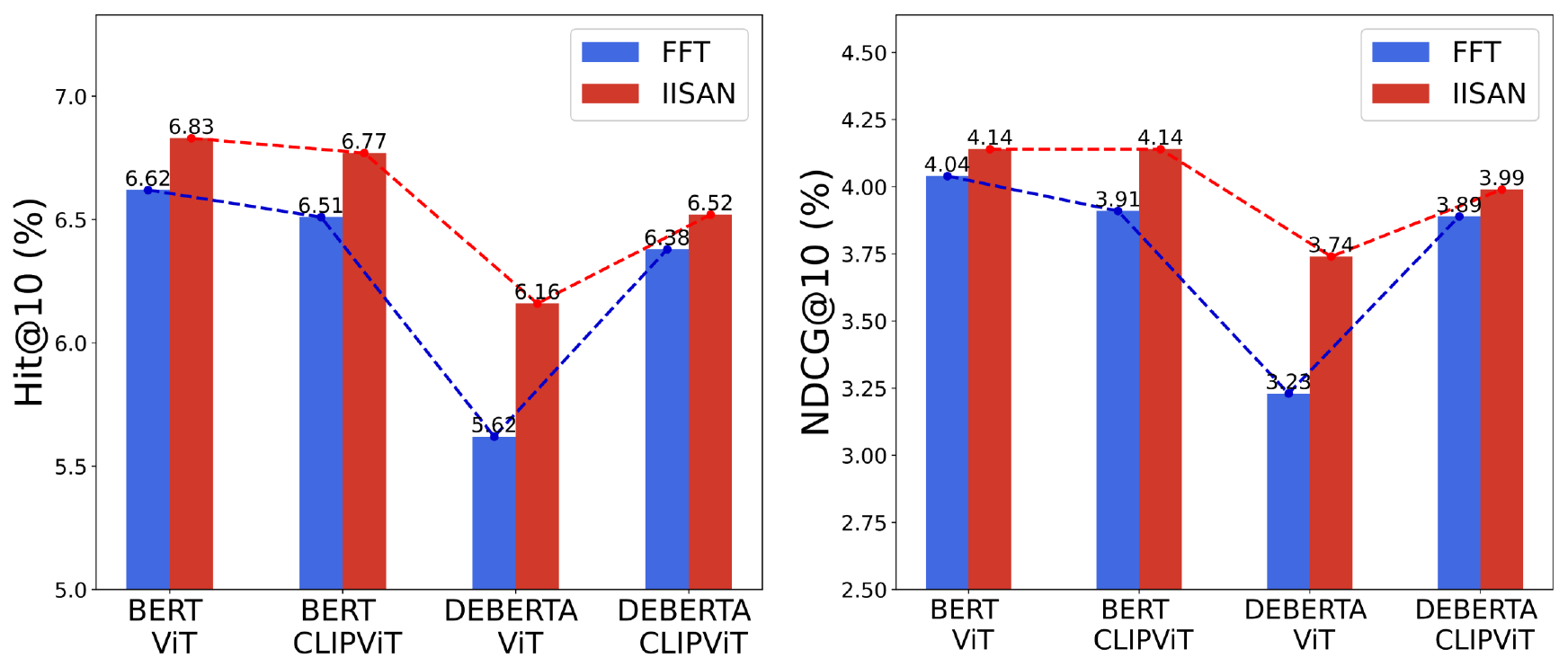}
   \vspace{-2.5em}
  \caption{Peformance comparisons between FFT and IISAN with different multimodal backbones on Scientific dataset. 
  }
    \label{fig:compare_fft_iisan_encoder} 
    \vspace{-0.2in}
\end{figure}

Table \ref{tab:eff_per} indicates that the proposed IISAN not only surpasses the performance of the FFT model on all datasets but only uses 22\% 
%\textcolor{red}{22 what? and the following sentence is incomplete}
 of the relative costs in terms of TPME. Furthermore, IISAN consistently outperforms all PEFT methods in both performance and practical efficiency under the same batch size (32). 
In particular, compared to the FFT model, using IISAN significantly improves the recommendation performances, e.g. by 3.07\%, 3.31\%, and 7.35\% of HR@10 and by 1.93\%, 3.57\%, and 6.91\% of NDCG@10 on Scientific, Instrument, and Office datasets, respectively. 
For model practical efficiency comparisons, IISAN achieves the best efficiency, whether compared with FFT or any EPEFT models. Without the caching strategy, IISAN can reduce the Training time and GPU memory by around 60\% and 82\%. Equipped with the caching strategy, the numbers even rise to 94\% and 93\%. While the EPEFT methods significantly reduce the number of trainable parameters, they do not substantially decrease the crucial aspects of model training, i.e. training time and GPU memory costs. 
According to the new balanced efficiency metric (TPME),  IISAN only uses about 22\% TPME cost of FFT, but the Adapter, LoRA and BitFit require around 71\%, 75\% and 70\% respectively.
Furthermore, IISAN with a caching strategy further reduces training costs and only needs 0.2\% TPME. 

\textbf{(Answer to RQ1) IISAN can achieve the most competitive performance with the highest efficiency and the TPME effectively uncovers the practical efficiency levels of each model. }
These experiments suggest that IISAN is a promising and efficient choice for representation learning tasks that involve multimodal item representation for recommendation.

\vspace{-0.1in}
\subsection{Robustness Evaluation (RQ2)}
To better understand the robustness of the proposed IISAN method, we evaluate it with four different combinations of state-of-the-art multimodal encoders based on~\cite{fu2024exploring}, which includes BERT+ViT, BERT+CLIPViT, DEBERTA+ViT, and DEBERTA+CLIPViT in Figure \ref{fig:compare_fft_iisan_encoder}. 
We observe that on the Scientific dataset, although the contribution of different backbones to multimodal recommendation is inconsistent, the proposed IISAN can always remain ahead of the FFT model. For example, when changing the text encoder and image encoder to  DEBERTA and CLIPViT respectively, IISAN still exceeds the FFT model by 0.12 in H@10 and 0.10 in NDCG@10.

\textbf{(Answer to RQ2) These results validate that IISAN maintains excellent robustness on different fundamental models.} This has reference implications for the model migration of IISAN.

\begin{table}
  \caption{Ablation study for IISAN on Scientific Dataset. IISAN mainly contains four key components: LayerDrop, Modality Gate, Intra- and Inter-modal towers. "-" represents TPME is not applicable for the frozen Backbone.}
  \vspace{-0.1in}
  \label{tab:abl_iisan}
  \renewcommand\tabcolsep{9.3pt}
  \renewcommand{\arraystretch}{0.8}
  \begin{tabular}{c | c  c c c}
    \hline
    \multirow{2}{*}{Method}&\multirow{2}{*}{HR@10}&\multirow{2}{*}{NDCG@10}&\multirow{2}{*}{TPME}\\
    &&&\\
    \hline
    -w/o LayerDrop&6.73 & 4.04&22.81\%\\
    \cdashline{2-4}
    -w/o Modality Gate&6.58 &3.89&  22.02\%\\
    \cdashline{2-4}
    Frozen Backbone &6.00 & 3.53&- \\
    -w/o Inter-modality&6.38 &3.89&21.31\%  \\
    -w/o Intra-modality&6.41 & 3.83&20.67\% \\
    \cdashline{2-4}
    IISAN&6.83 & 4.14 &22.34\%\\
  \hline
\end{tabular}
\vspace{-0.15in}
\end{table}

\begin{table}
  \caption{LayerDrop in IISAN on Scientific Dataset. The number of blocks in the Method column represents that keep this number of blocks and drop the others.}
    \vspace{-0.1in}
  \label{tab:layerdrop}\renewcommand{\arraystretch}{0.7}
  \begin{tabular}{c | c | c  c c}
    \hline
    \multirow{2}{*}{Method} & \multirow{2}{*}{Number of blocks} & \multirow{2}{*}{HR@10} & \multirow{2}{*}{NDCG@10} & \multirow{2}{*}{TPME}\\
    & & & & \\
    \hline
    \multirow{5}{*}{IISAN} & 2 blocks & 6.46 & 3.88 & 22.05\% \\
    & 3 blocks & 6.79 & 4.12 & 22.06\% \\
    & 4 blocks & 6.57 & 3.98 & 22.10\% \\
    & \textbf{6 blocks} & \textbf{6.83} & \textbf{4.14} & 22.34\% \\
    & 12 blocks & 6.73 & 4.04 & 22.81\% \\
  \hline
  \end{tabular}
  \vspace{-0.2in}
\end{table}

\subsection{Ablation Study (RQ3)}
\label{sec:rq3}
To comprehensively answer the RQ3, we perform extensive ablation studies. We demonstrate how each component affects the overall performance and efficiency of IISAN. Note that the decoupled structure and caching strategy are the main contributors to the efficiency gains as described in Section \ref{sec:theory}. Table \ref{tab:abl_iisan} presents component ablation studies focusing on the various modules within IISAN, including (1) Modality selection, (2)  LayerDrop strategies, (3) Gated fusion, and (4) Implementation of SANB.

\textbf{(1) Modality selection.} Table \ref{tab:abl_iisan} highlights that although employing separate intra-modal SAN (Line 4) or inter-modal SAN (Line 5) can slightly improve efficiency, it will clearly reduce the recommendation effects. However, compared the pre-trained backbones (Line 3) with frozen layers (only training the sequential encoder), both intra-modal SAN and inter-modal SAN can significantly improve the performances and perform best when used simultaneously. It reflects that intra- or inter-modal SAN can improve feature adaptation within each modality and inter-modal feature interaction adaptation. 

\textbf{(2) LayerDrop strategies.} IISAN's layers exhibit a degree of redundancy, allowing the utilization of LayerDrop to enhance performance effectively. 
First, when we remove the LayerDrop, both HR@10 and NDCG@10 decrease absolutely by 0.1\% and the cost slightly rises by 0.5\% on TPME, which demonstrates the effectiveness of LayerDrop. 
Second, we study the effect of different LayerDrop strategies as described in Table \ref{tab:layerdrop}. We adopt the even number $(2,4,6,...12)$ transformer-blocks (6 blocks), which skip the odd number layers. It achieves the best performance-efficiency balance. 

\textbf{(3) Gated fusion.} To deepen our understanding of the differences in information usage between textual and visual modalities in our research, we analyzed the gate weights at the optimal checkpoint. These gate values, ranging from 0 to 1, imply that values above 0.5 denote a stronger focus on a specific modality as discussed in Section \ref{sec:iisan}. The data indicate that in the realm of multimodal recommendations, the weights assigned to visual modality gates consistently ranged from 0.2 to 0.4. This trend suggests a predominant reliance on textual modality within our multimodal approach.

\textbf{(4) Implementation of SANB.} Table \ref{tab:sanb} demonstrates the notable superiority of the classic adapter block in terms of performance compared with other more recent PHM (Parameterized Hypercomplex Multiplication) block~\cite{karimi2021compacter} and Low-Rank adapter blocks~\cite{yin20231}. Their performance, as indicated by HR@10, is significantly lower than 6.62 (FFT), suggesting that their adaptation is unsuccessful. This phenomenon is corroborated by findings in \cite{fu2024exploring} where PHM shows a performance drop compared to traditional adapters.
Notably, both PHM and LowRank models, despite having merely half the trainable parameters of an adapter block, exhibit comparable training time and GPU memory, resulting in similar TPME. 

\textbf{(Answer to RQ3) In this section, we conclude with three key findings of the IISAN's components: (1) Optimal performance is achieved through the placement of both intra- and inter-modal Self-Attention Networks (SANs). 
(2) The best efficiency and performance balance is achieved by dropping half of the SANBs in the Intra- and inter-modal SANs. 
(3) Text-image interaction is effective, but text modality plays a more crucial role in recommendation, where inter-modal SAN can effectively maintain the dominance of text modality and integrate image information.}
\begin{table}
  \caption{Implementation of SANB (Side Adapted Network Block) on Scientific Dataset. We get the TPME of uncached  approach. }
  \vspace{-0.15in}
  \label{tab:sanb}
  \renewcommand{\arraystretch}{0.9}
  \begin{tabular}{c c | c  c c}
    \hline
     \multirow{2}{*}{Model}& \multirow{2}{*}{Implementation}&\multirow{2}{*}{HR@10}&\multirow{2}{*}{NDCG@10} & \multirow{2}{*}{TPME}\\
    & &&&\\
    \hline
    \multirow{3}{*}{SANB}  & Adapter Block\cite{houlsby2019parameter}&6.73 &4.04 &22.81\%\\
      &PHM\cite{karimi2021compacter}&5.95 &3.59 &22.34\% \\
      &LowRank\cite{yin20231}&4.79 & 2.83 & 22.47\%\\
  \hline
\end{tabular}
\vspace{-0.3in}
\end{table}

\subsection{Multimodality vs. Unimodality (RQ4)}
\label{sec:rq4}
The concept of PEFT in modality-based recommendation research is relatively nascent, with limited literature directly comparing multimodal and unimodal scenarios. Driven by this curiosity, we conduct experiments on the performance of FFT and various PEFTs across different modality scenarios.

Table \ref{tab:uni_vs_multi} presents comparisons between unimodal and multimodal scenarios. 
First, the PEFT models (such as Adapter, LoRA, and BitFit) exhibit performance comparable to FFT in text-based and multimodal scenarios, but poor image-based performance.
Additionally, the text-based adapter significantly outperforms FFT, indicating its high suitability for text-based recommendation. 
This may be caused by the inherent bias in the modal transfer  from the foundation models in multimodal recommendation tasks where images are more difficult. 
This observation is corroborated by our analysis of information usage in the gated fusion (in Section \ref{sec:rq3}), where text modality is dominant in the inter-modal interaction.
Second, each adaptation approach based on multimodal representations demonstrates superior performance, underscoring the importance of integrating multiple modalities. In particular, IISAN achieves the best results due to the novel intra- and inter-modal adapted networks.

\begin{table}
  \caption{Multimodality vs. Unimodality.}
  \label{tab:uni_vs_multi}
  \vspace{-0.15in}
  \renewcommand\tabcolsep{8.5pt}
  \renewcommand{\arraystretch}{0.8}
  \begin{tabular}{c|c|cc}
    \hline
       \multirow{2}{*}{Modality} & \multirow{2}{*}{Method} & \multirow{2}{*}{HR@10} & \multirow{2}{*}{NDCG@10} \\
    & & & \\
    \hline
    \multirow{4}{*}{Image}          & FFT&5.81 & 3.47 \\
                                   & Adapter\cite{houlsby2019parameter} &5.47 & 3.18 \\
                                   & LoRA\cite{hu2021lora} & 5.49 &3.34 \\
                                   & BitFit\cite{zaken2021bitfit} &4.97 &3.03 \\
    \hline
    \multirow{4}{*}{Text} & FFT &5.80 &3.44 \\
                                   & Adapter\cite{houlsby2019parameter} &6.55 & 3.88 \\
                                   & LoRA\cite{hu2021lora} &6.14 & 3.61 \\
                                   & BitFit\cite{zaken2021bitfit} &5.92 & 3.56 \\
    \hline
    \multirow{4}{*}{Multimodality} & FFT &6.62 & 4.06 \\
                                   & Adapter\cite{houlsby2019parameter} &6.61 &3.91 \\
                                   & LoRA\cite{hu2021lora} &6.62 &4.09 \\
                                   & BitFit\cite{zaken2021bitfit} &6.37 & 3.76 \\
                                   \cline{2-4}
                                   & \textbf{IISAN (ours)} &\textbf{6.83} & \textbf{4.14} \\
    \hline
  \end{tabular}
  \vspace{-0.25in}
\end{table}

\textbf{(Answer to RQ4) The findings suggest that multimodality is more advantageous than relying solely on unimodality.} 

\vspace{-0.1in}
\section{Related Work}
\textbf{Modality-based Sequential Recommendation (MoRec)}.
 The recommendation with various modality information has increasingly fascinated the Recommender System community \cite{wu2021mm,sun2020multi,yuan2023go,wang2022transrec,li2023exploring,zhang2023ninerec,wei2023multi,wang2023missrec,cheng2023image,ni2023content,liu2023id,qu2023thoroughly,hu2024lightweight,liu2024once}. They deploy the large-scale pre-trained foundation models  \cite {devlin2018bert,he2021debertav3,he2020deberta,brown2020language,He2016deep,radford2021learning} from NLP and CV
\cite{dosovitskiy2020image,liu2021swin}, and CLIP \cite{radford2021learning} 
to encode texts and images. The sequential encoder stays unchanged as the traditional sequential architectures, including SASRec \cite{kang2018self},  GRU4Rec \cite{hidasi2015session}, and BERT4Rec \cite{sun2019bert4rec}, etc. In addition, IDA-SR \cite{mu2022id}, UniSRec \cite{hou2022towards}, and VQ-Rec \cite{hou2022learning} realized MoRec by learning item representation from the foundation model in NLP. In terms of image representation in RS,
\cite{wei2019mmgcn} and~\cite{meng2020heterogeneous} fed the RS model with image features extracted from the ResNet-50~\cite{He2016deep}.  
On the other hand,
\cite{elsayed2022end,yuan2023go,ni2023content,li2023multi} 
 have collectively demonstrated that the MoRec framework, when applied through end-to-end learning, significantly enhances performance over previous methods that rely on offline feature extractors. Specifically,  \cite{li2023multi, li2024empirical} highlighted that end-to-end training, which integrates both image and text modalities, considerably surpasses systems based solely on a single modality. 
 
 However, a significant limitation observed in these end-to-end research studies is their continued reliance on full fine-tuning of large-scale multimodal encoders. This approach often results in a lack of efficiency.

\noindent\textbf{Parameter-efficient Fine-tuning (PEFT)}. In the domains of NLP and CV, significant efforts are being devoted to investigating PEFT methods. These approaches aim to address the challenge of managing the extensive number of trainable parameters present in large-scale pre-trained models. Foundational works in this area include Adapter~\cite{houlsby2019parameter}, LoRA~\cite{hu2021lora}, and BitFit~\cite{zaken2021bitfit}. Building on this paradigm, a variety of alternative strategies have been proposed, as evidenced by studies such as \cite{karimi2021compacter,pfeiffer2020adapterhub,wang2020k,zhang2021tip,chen2022vision,he2022parameter,gao2023llama,yang2023tackling}. Nevertheless, these methods predominantly employ Embedded PEFT (EPEFT) and concentrate chiefly on parameter efficiency, rather than practical efficiency factors like training speed and GPU memory consumption.
%, which are detailed in Section \ref{sec:intro}
To address these limitations, \cite{cai2020tinytl} introduced the concept of "Reduce Memory, Not Parameters", suggesting a reduction in activations.   LST \cite{sung2022lst} introduces a memory efficiency approach that tries to decouple PEFT modules from the encoder-decoder T5 \cite{raffel2020exploring}. This emerging trend in NLP and CV research is increasingly pivoting from the earlier focus on EPEFT to DPEFT, as exemplified by recent studies \cite{xu2023side,lin2023vision,xu2023san}.

For the usage of PEFT in the realm of modality-based sequential recommender systems (RS), significant progress has been demonstrated by M6-Rec~\cite{cui2022m6}, Tallrec~\cite{bao2023tallrec}, and SSNA~\cite{peng2023towards}. These studies highlight the PEFT approaches can be adopted to achieve comparable performance to traditional fine-tuning methods. Additionally, \cite{fu2024exploring} provides an empirical study of adapters in MoRec.  Although many current methods in the field continue to utilize the conventional EPEFT, they often overlook aspects of practical efficiency. The adoption of DPEFT in RS remains limited, with only a few exceptions, such as a concurrent preprint~\cite{peng2023towards}. Furthermore, these methods mainly concentrate on single-modality analysis, missing the opportunity to exploit the rich multimodal information available in RS. In contrast, VIP5~\cite{geng2023vip5} explores the multimodal recommendation using adapters within the P5 backbone. However, its implementation primarily utilizes adapters for text encoders, leaving the image encoder unchanged and dedicated solely to feature extraction. This approach differs from our focus, as adapting the image encoder incurs higher costs compared to the text encoder.

To the best of our knowledge, little research investigated practical efficiency issues of using DPEFT approaches for multimodal representation adaptation for recommendation tasks. %Multimodal scenarios can be fairly different from only using one modality as described in Section \ref{sec:rq3}. 
Utilizing multimodal information can certainly boost recommendation performance \cite{li2023multi}, while further increasing the practical efficiency issues. Therefore, the research of studying DPEFT approaches to address the real efficiency problems in this field is a novel and urgent direction. 

\vspace{-0.2in}
\section{Conclusion and Future Work}
In this study, we present a novel Decoupled-PEFT architecture, named IISAN, for recommendation task adaptation of pre-trained large-scale multimodal fundamental models.
IISAN leverages advantages inherent in DPEFT to separate the trainable intra- and inter-modal adaption networks from the multimodal backbones, thus minimizing the computational graph and allowing for caching strategies. 
These allow IISAN to optimize in terms of model practical efficiency. 
Besides, the new intra- and inter-modal SANs in IISAN achieve comparable performance to the full fine-tuning model by combining both intra- and inter-modal information adaptive interactions. 
In addition, we introduce a balanced efficiency metric, --TPME--, to evaluate the multi-faceted practical model efficiency between different methods. 
Finally, experimental results on three recommendation datasets have shown the superiority of both efficiency and effectiveness of IISAN. And the efficiency analysis also proves its high efficiency. 

Future work includes the exploration of more potential applications, such as multimodal retrieval\cite{ge2023cross,ge20243shnet} and visual question answering (VQA) \cite{zhou2021trar}, etc., by the IISAN paradigm. 
Moreover, many more modality representations can be applied by the novel intra- and inter-modal SANs, e.g. image, text audio, etc., to further adapt the multimodal real-world scenarios.

\begin{acks}
This research was supported in part by China Scholarship Council (CSC) from the Ministry of Education of China (No. 202308330014).
\end{acks}
\bibliographystyle{ACM-Reference-Format}
\balance
\bibliography{sample-base}

%%
%% If your work has an appendix, this is the place to put it.
\appendix
\clearpage

\end{document}